\documentclass[runningheads, a4paper]{llncs}

\usepackage{cite}
\usepackage{url}
\usepackage{tabularx}
\usepackage[dvipdfmx]{graphicx}
\usepackage{amsmath, amssymb}
\usepackage{algorithm, algpseudocode}

\newcommand\pfun{\mathrel{\ooalign{\hfil$\mapstochar\mkern5mu$\hfil\cr$\to$\cr}}}
\newcommand{\reqas}[1]{\mathrm{req\_as}(#1)}
\newcommand{\reqatext}{req\_a}
\newcommand{\reqa}[1]{\mathrm{\reqatext}(#1)}
\newcommand{\reqp}[1]{\mathrm{req\_p}(#1)}
\newcommand{\introa}[2]{\mathrm{int\_a}_#1(#2)}
\newcommand{\introp}[2]{\mathrm{int\_p}_#1(#2)}
\newcommand{\appears}[1]{\mathrm{appear}(#1)}
\newcommand{\powerset}[1]{2^{#1}}
\begin{document}

 \mainmatter

        \title{Towards Refinement Strategy Planning for Event-B}
 \titlerunning{Towards Refinement Strategy Planning for Event-B}
 
        \author{Tsutomu Kobayashi\inst{1} \and Shinichi Honiden\inst{1}\inst{2}}
 \authorrunning{Tsutomu Kobayashi and Shinichi Honiden}
 
 \urldef{\mails}\path|{t-kobayashi, honiden}@nii.ac.jp|

 \institute{The University of Tokyo, Japan \and National Institute of Informatics, Japan \\
 \mails \\
 \url{}}
 
 \maketitle

 \begin{abstract}
  Event-B is a formal approach oriented to system modeling and analysis. It supports refinement mechanism that enables stepwise modeling and verification of a system. By using refinement, the complexity of verification can be spread and mitigated. In common development using Event-B, a specification written in a natural language is examined before modeling in order to plan the modeling and refinement strategy. After that, starting from a simple abstract model, concrete models in several different abstraction levels are constructed by gradually introducing complex structures and concepts. Although users of Event-B have to plan how to abstract the specification for the construction of each model, guidelines for such a planning have not been suggested. Specifically, some elements in a model often require that other elements are included in the model because of semantics constraints of Event-B. As such requirements introduces many elements at once, non-experts of Event-B often make refinement rough though rough refinement does not mitigate the complexity of verification well. In response to the problem, a method is proposed to plan what models are constructed in each abstraction level. The method calculates plans that mitigate the complexity well considering the semantics constraints of Event-B and the relationships between elements in a system.
  \keywords{Formal Methods, Refinement, Event-B, Specifications, Abstraction}
 \end{abstract}

 \section{Introduction}
 Event-B\cite{Event-B-book} is a formal approach oriented to system-level analysis and modeling. Event-B users specify models in a notation based on set theory and first order logic and check the model by proving.

 The most notable feature of Event-B is the support for refinement mechanism. In a refinement process, users first construct a simple and highly abstract model. After checking the consistency in the abstract model, more complex model is constructed and the consistency between the abstract model and the complex model is checked. Usually complex models are constructed by introducing new aspects and properties of a system. Users gradually construct concrete models by repeating this step. Refinement enables users to construct a complex model more simply rather than to construct the complex model at once. Therefore, the burden of a model construction is mitigated by Event-B.

 As Event-B is an effective method, it has been attracting more and more attentions from the industry. For example, many companies in Japan are interested in formal methods including Event-B, while technical and administrative guidelines are constructed through cooperation of a national institute and software vendors \cite{DSF}.

 Refinement enables users to spread the complexity of modeling over some steps. However, it is necessary to properly define how elements are gradually introduced, mitigating the complexity while complying with semantics constraints of Event-B. In this paper, we propose a method that considers the constraints and relationships between elements of a system and plans what models should be constructed for an effective refinement. Thus the method enables ordinary users to leverage the refinement mechanism in a simple way.

 The remainder of this paper is organized as follows. Section~\ref{151914_6Oct12} describes the problem and the cause of the problem. Section~\ref{134113_11Oct12} describes the proposed method together with exemplification on an example. Section~\ref{152306_10Oct12} shows related work as well as future work, before concluding remarks in Section~\ref{134138_11Oct12}.

 \section{Problem and Approach}
  \label{151914_6Oct12}
  In usual modeling in Event-B, users first read the specification of the system written in a natural language. Then models in several abstraction levels are constructed gradually. Event-B supports checking constructed models but does not guide modeling explicitly. Thus users have to plan what models are constructed in each abstraction level. Specification of a system is composed of a set of statements about property of the system. We call such statements \textit{artifact}s. For example, a specification of a library management system may include an artifact ``There are no loaned books in the open stack.''. Usually, Event-B models include invariants that correspond to a subset of artifacts of the system. Thus, users have to plan which subset of artifacts of the system should be reflected to models of each abstraction level. When constructing a concrete model, new artifacts are added to artifacts of the abstract model. Therefore users need to plan which artifacts are introduced to each abstraction level.
 
  In Event-B models, artifacts are expressed as invariants using the formal language of Event-B. Thus, in order to express an artifact, it is necessary to introduce elements (e.g. career sets, constants, variables, and events) that correspond to terms appeared in the artifact. We call such terms \textit{phenomena}. For instance, an artifact ``There are no loaned books in the open stack.'' can be expressed like ``$openstack \,\cap\, \mathrm{dom}(loan\_state) = \emptyset$'' (Where $openstack \subseteq books$ and $loan\_state \in books \pfun members$) only when $openstack$ and $loan\_state$ are already introduced to the model.
 
  Such constraints on the introduction of phenomena exist not only between an artifact and phenomena but also between a phenomenon and phenomena. That is, in some cases, an introduction of a phenomenon requires an introduction of some other phenomena. The causes of such constraints on introduction include the following two facts.
 
  Firstly, all variables and constants in an Event-B model have to be typed as a primitive type (a built-in data type or an element of a career set as a user-defined type) or a pair that is recursively built from primitive types. A primitive type is atomic and not expressible by using other primitive types. For example, if a variable $var$ is typed as $S$ by using a career set $S$ in an abstract model and typed as $T \rightarrow \mathbb{N}$ by using a career set $T$ and a built-in data type $\mathbb{N}$ in a concrete model, a type error will occur. Thus, any typing statement of a phenomenon should not be changed through refinement. Therefore, a phenomenon corresponds to a variable or a constant can be introduced only when phenomena that are necessary to type the variable or constant are already introduced. For instance, in order to introduce a phenomenon ``loan state'' and express its type as $loan\_state \in books \pfun members$, the introduction of career sets $books$ and $members$ is needed. Thus, it is also required to introduce phenomena that corresponds to career sets $books$ and $members$.
 
  Secondly, Event-B has several criteria for consistency between an abstract model and a concrete model. In modeling in Event-B, users can confirm consistency of models by proving them (proof obligations). There is a proof obligation named EQL, which requires that if a variable is included in both an abstract model and a concrete model there must not be a state transition such that it is included in the concrete model but not included in the abstract model. In order to make a refinement consistent EQL proof obligation must hold. Therefore a phenomenon corresponds to a variable can be introduced in a model only when state transitions that change the value of the variable are already introduced. For example, the introduction of a variable that corresponds to a phenomenon ``loan state'' requires an introduction of all state transitions included in the behavior of the system (e.g. ``loaning a book from the open stack'', ``returning a book'', ``loaning a reserved book'').
 
  For these reasons, an introduction of an artifact to a model requires introductions of 1) phenomena that appear in the artifact and 2) phenomena that are required by 1). Let $A = (A_i)_{i = 0, 1, \cdots} (A_0 = \emptyset)$ be a sequence of sets of artifacts reflected to the $n$th model, $\reqa{a}$ be the phenomena required by an introduction of an artifact $a$, and $\reqas{A_i} = \bigcup_{a \in A_i} \reqa{a}$ for a set of artifacts $A_i$. When an artifact is introduced, phenomena that are not introduced yet but required for an introduction of the artifact are also introduced. Then, let $\introp{i}{A} = \reqas{A_i} \setminus \reqas{A_{i-1}} \; (1 < i)$\footnote{In this paper, the relative complement of a set $T$ in a set $S$ is denoted by $S \setminus T$.}. It denotes newly installed phenomena to the $i$th model in a sequence of artifacts sets $(A_i)_{i = 0, 1, \cdots}$.

\begin{figure}[h]
   \begin{center}
    \includegraphics[width=\textwidth]{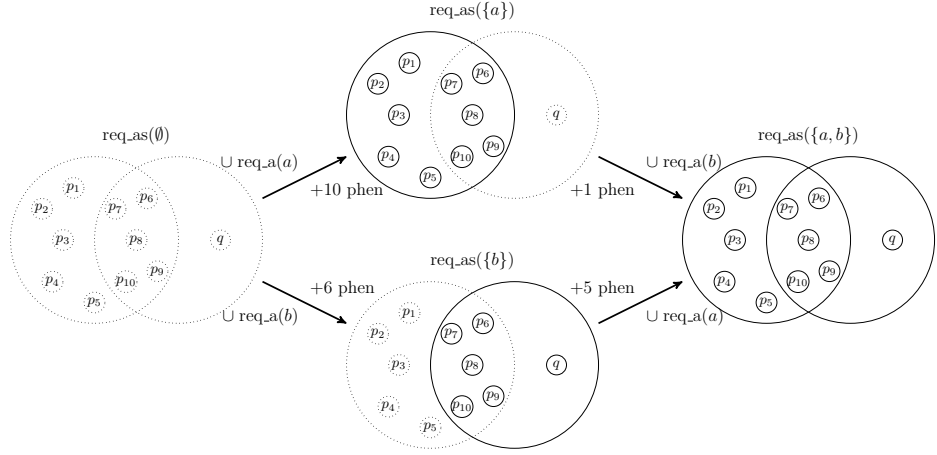}
   \end{center}
   \caption{Difference between Two Introduction Orders}
   \label{164014_15Oct12}
  \end{figure}

  Consider the introduction order of artifacts $a$ and $b$ such that $\reqa{a} = \{p_1, \cdots, p_{10}\}$, $\reqa{b} = \{p_6, \cdots, p_{10}, q\}$ to an empty model (Figure~\ref{164014_15Oct12}). For $A_1 = \{a\}, A_2 = \{a, b\}$, the newly introduced phenomena will be as follows: $\introp{1}{A} = \reqa{a} = \{p_1, \cdots, p_{10}\}$, $\introp{2}{A} = \reqa{b} \setminus \reqa{a} = \{q\}$. In contrast, For $B_1 = \{b\}, B_2 = \{a, b\}$, the newly introduced phenomena will be as follows: $\introp{1}{B} = \reqa{b} = \{p_6, \cdots, p_{10}, q\}$, $\introp{2}{B} = \reqa{a} \setminus \reqa{b} = \{p_1, \cdots, p_5\}$.

  Therefore, $(|\introp{1}{A}|$, $|\introp{2}{A}|)$ $= (10, 1)$, whereas $(|\introp{1}{B}|$, \linebreak $|\introp{2}{B}|)$ $= (6, 5)$. Thus, the number of newly introduced phenomena in each model will vary depending on the introduction order of artifacts. As refinement is a mechanism to mitigate the complexity of modeling by spreading the introduction of phenomena over some steps, how much the numbers of newly introduced phenomena are dispersed is useful information for users to plan refinement. Thus, we define the \textit{effectiveness} of an introduction order of artifacts as in Definition~\ref{def:eff}. For instance, the order $(a, b)$ is more effective than $(b, a)$ in the above example. Moreover, an order $(A_i)_{i = 0, \cdots ,3}$ such that $(num_{A,i})_{i=1,2,3} = (3,1,2)$ is more effective than an order $(B_i)_{i = 0, \cdots ,3}$ such that $(num_{B,i})_{i=1,2,3} = (0,3,3)$ since $(snum_{A,i})_{i=1,2,3} = (3,2,1)$ is smaller than $(snum_{B,i})_{i=1,2,3} = (3,3,0)$ in lexicographical order.
  \begin{definition}\label{def:eff}
   Let a sequence $(num_{A,i})_{i = 1, \cdots, |A|}$ be a history of the number of phenomena newly introduced in each refinement (i.e. $num_{A,i} = |\introp{i}{A}|$) and a sequence $(snum_{A,i})_{i = 1, \cdots, |A|}$ be a sorted permutation of $(num_{A,i})_{i = 1, \cdots, |A|}$ in descending order. Then, for a sequence $(A_i)_{i = 0, \cdots, |A|}$ and $(B_i)_{i = 0, \cdots, |B|}$ such that $\reqas{A_{|A|}} = \reqas{B_{|B|}}$, $(A_i)_{i = 0, \cdots, |A|}$ is called \textit{more effective} than $(B_i)_{i = 0, \cdots, |B|}$, if $(snum_{A,i})_{i = 1, \cdots, |A|}$ is smaller than $(snum_{B,i})_{i = 1, \cdots, |B|}$ in lexicographical order.
  \end{definition}
  As in the above example, users should plan an introduction order of artifacts so that the refinement is effective. However, for this planning, users have to grasp and compare the constraints on introduction between phenomena over multiple steps. The constraints are too complex for users to analyze in their heads. Therefore, Event-B users (especially beginners) have to repeat trial and error processes during modeling many times. Thus, though refinement is a powerful mechanism, it is not so easy to use refinement in realistic situations.

 \section{Method}
 \label{134113_11Oct12}
  \subsection{Derivation of Required Phenomena}
  As we viewed in Section~\ref{151914_6Oct12}, the effectiveness of refinement depends on the sets of phenomena required by artifacts. The phenomena required by an artifact depend on types and state transitions related to phenomena that appear in the artifact. In the proposed method, constraints on introduction between phenomena related to types and state transitions are assumed as the input. The output of the method is orders that maximize the effectiveness of refinement.

  To illustrate the method, construction of a model of a library management system is described. The artifacts of the system are as shown in Table~\ref{070847_7Oct12} and the events of the system are as shown in Table~\ref{033418_10Oct12}. Events represent behavior of the system. An event can cause multiple state transitions. 

  \begin{table}[t]
   \caption{Artifacts of Library Management System}
   \label{070847_7Oct12}
   \begin{tabularx}{\linewidth}{c X X}
    \hline
     & Artifact & Phenomena Appeared in the Artifact \\
    \hline
    $a$ & ``Loan is done only for members'' & loan state, members \\
    $b$ & ``Books on loan are not in the open stack'' & loan state, books, open stack state \\
    $c$ & ``No reserved books are in the open stack'' & reservation state, books, open stack state \\
    \hline
   \end{tabularx}
  \end{table}
 
  \begin{table}[t]
   \caption{Events of Library Management System}
   \label{033418_10Oct12}
   \begin{tabularx}{\linewidth}{l l X}
    \hline
    & Event & Caused State Transitions \\
    \hline
    p1 & Loaning a reserved books & Remove one from reservation state, \\
    & & Add one to loan state \\
    p2 & Returning a book & Remove one from loan state \\
    p3 & Loaning a book from the open stack & Remove one from open stack state,\\
    & & Add one to loan state \\
    \hline
   \end{tabularx}
  \end{table}
 
  \begin{figure}[t]
   \begin{center}
    \includegraphics[width=.9\textwidth]{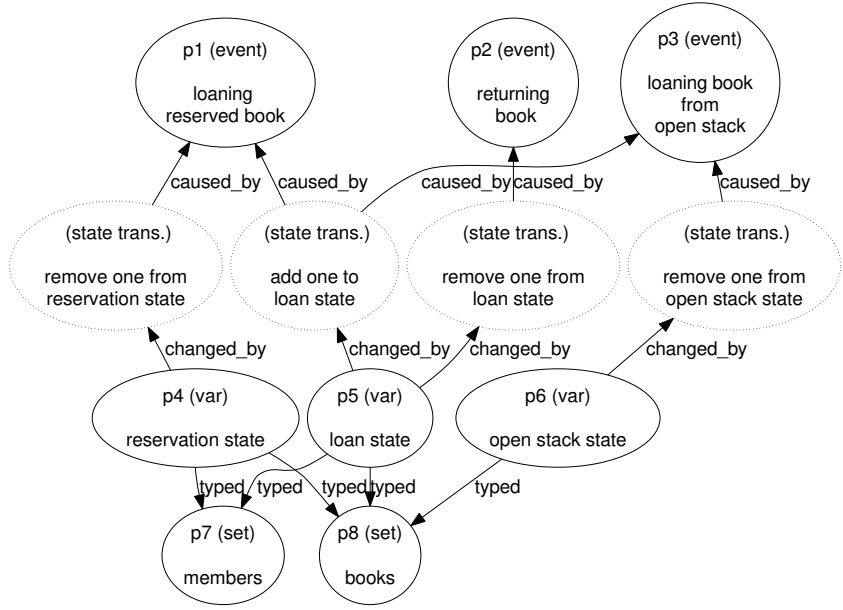} \\
    \end{center}
   \caption{Constraints on Introduction between Phenomena in Library Management System}
   \label{071252_7Oct12}
  \end{figure}

  Phenomena can be classified into four kinds according to what kind of element in an Event-B model corresponds to the phenomenon. Phenomenon corresponds to a career set, a constant, a variable, or an event.
  
  Let $P_{\mathrm{S}}$, $P_{\mathrm{C}}$, $P_{\mathrm{V}}$, and $P_{\mathrm{E}}$ be the set of phenomena that correspond to career sets, constants, variables, and events, respectively. Let $P = P_{\mathrm{S}} \cup P_{\mathrm{C}} \cup P_{\mathrm{V}} \cup P_{\mathrm{E}}$ and $T$ be the set of state transitions.

  Let $typed: P \rightarrow \powerset{P_{\mathrm{S}}}$ be a set of career sets required for typing a constant or a variable, $changed\_by: P \rightarrow \powerset{T}$ be a set of state transitions that change the value of a variable, and $caused\_by: T \rightarrow \powerset{P_{\mathrm{E}}}$ be a set of events that causes a state transition.

  The proposed method takes these three functions as the input from the users. For example, constraints on introduction between phenomena in the library management system can be depicted as in Figure~\ref{071252_7Oct12}.

  The set of career sets required for typing a phenomenon can be derived by tracing $typed$ edges. The set of events that change a phenomenon can be derived by tracing $changed\_by$ edges and $caused\_by$ edges. Therefore, let $\reqp{p}$ be the phenomena required by an introduction of a phenomenon $p$, then $\reqp{p}$ can be derived by the input. That is,
  \begin{align}
   \reqp{p} = typed(p) \,\cup\, \bigcup_{t \in changed\_by(p)} caused\_by(t)\label{111208_10Oct12} \enspace .
  \end{align}
  Thus, $\reqa{a}$ can be derived by the input. That is,
  \begin{align}
   \reqa{a} = \appears{a} \,\cup\, \bigcup_{p \in \appears{a}} \reqp{p}\label{111054_10Oct12} \enspace .
  \end{align}
  Where $\appears{a}$ denotes the phenomena that directly appear in an artifact $a$. By constraints on introduction depicted in Figure~\ref{071252_7Oct12} and Equations (\ref{111208_10Oct12}) and (\ref{111054_10Oct12}), $\mathrm{\reqatext}$ of each artifact in the library management system is derived as follows:

  \begin{center}
   \begin{tabular}[t]{cccccccccc}
    $\reqa{a} = \{$ & p1, & p2, & p3, &     & p5, &     & p7, & p8 & $\}$ \\
    $\reqa{b} = \{$ & p1, & p2, & p3, &     & p5, & p6, & p7, & p8 & $\}$ \\
    $\reqa{c} = \{$ & p1, &     & p3, & p4, &     & p6, & p7, & p8 & $\}$ \\
   \end{tabular}
  \end{center}

  \subsection{Search for the Best Introduction Orders of Artifacts}
  Let $\introa{i}{A} = A_i \setminus A_{i-1} \; (1 < i)$. It denotes newly introduced artifacts to the $i$th model. We assume that only one artifact is introduced through one refinement step. Then, $|\introa{i}{A}| = 1 \; (1 < i)$ and $\reqas{A_i}= \reqa{a} \,\cup\, \reqas{A_{i-1}} \; (1 < i)$ where $a = |\introa{i}{A}|$.

  In the proposed method, orders that correspond to sequences $(A_i)_{i = 0, \cdots, |A|}$ that maximized the effectiveness of the refinement are obtained by Algorithm~\ref{065635_9Oct12}. All orders of artifacts introduction ($(\introa{i}{A})_{i=1,2,3}$) and the numbers of newly introduced phenomena ($(|\introp{i}{A}|)_{i=1,2,3}$) in each refinement for the library management system is as shown in Table~\ref{033550_10Oct12}. In this case, the result of the algorithm represents the order $(\{a\}, \{b\}, \{c\})$.

  The method uses breadth first search with pruning as shown in the Algorithm~\ref{065635_9Oct12}. A node of the search tree represents an introduction order of artifacts. A structure that represents a node is composed of $as$ that represents the history of artifacts introduction, $ps$ that represents the set of phenomena introduced so far, $nums$ that represents the history of the number of introduced phenomena in each step, $max$ that represents the maximum of $nums$, and $rest$ that represents the number of phenomena not introduced yet.

  The function \textsc{CertainlyBetter} (Line~\ref{algfun:better_than_begin}--\ref{algfun:better_than_end}) checks whether an introduction order of artifacts is certainly effective than the other order. This function is used for pruning (Line~\ref{alg:pruning1}, \ref{alg:pruning2}). The number of phenomena introduced in a later refinement is at most the number of phenomena not introduced yet \\ ($maybe\_better.rest$) since $\reqas{A_i} \subseteq \reqas{A_{|A|}}$ for all $i$. Therefore, an order is certainly better if the maximum number of introduced phenomena in each step of the order is at most less than the current maximum of the other order (Line~\ref{alg:true_begin}--\ref{alg:true_end}). If both maximums are equal, the algorithm retries the checking on orders without the artifact corresponds to the maximum number (Line~\ref{alg:reduce_and_recurse_begin}--\ref{alg:reduce_and_recurse_end}).

  \begin{algorithm}
   \caption{Search the Best Introduction Orders of Artifacts}
   \label{065635_9Oct12}
   \begin{center}
    \begin{algorithmic}[1]
    \Function{CertainlyBetter}{$maybe\_better, maybe\_worse$}
     \label{algfun:better_than_begin}
     \If{$((maybe\_better.nums = \{\}) \vee (maybe\_worse.nums = \{\}))$}
      \State \Return $false$
      \Comment{Not sure whether $maybe\_better$ is better}
     \ElsIf{$\{max(\{maybe\_better.max, maybe\_better.rest\}) <$ \\ $\quad \quad \quad \quad \quad \; maybe\_worse.max)$}\label{alg:true_begin}
      \State \Return $true$\label{alg:true_end}
      \Comment{$maybe\_better$ is certainly better}
     \ElsIf{$maybe\_better.max = maybe\_worse.max$}\label{alg:reduce_and_recurse_begin}
      \State new $mb\_reduced, mw\_reduced$
      \State $mb\_reduced.nums \leftarrow maybe\_better.nums \setminus \{maybe\_better.max\}$
      \State $mw\_reduced.nums \leftarrow maybe\_worse.nums \setminus \{maybe\_worse.max\}$
      \State $mb\_reduced.max \leftarrow max(mb\_reduced.nums)$
      \State $mw\_reduced.max \leftarrow max(mw\_reduced.nums)$
      \State $mb\_reduced.rest \leftarrow maybe\_better.rest$
      \State $mw\_reduced.rest \leftarrow maybe\_worse.rest$
      \State \Return \Call{CertainlyBetter}{$mb\_reduced, mw\_reduced$}\label{alg:reduce_and_recurse_end}
     \Else
      \State \Return $false$
      \Comment{Not sure whether $maybe\_better$ is better}
     \EndIf
    \EndFunction\label{algfun:better_than_end}
    \Statex
    \Function{SearchBestOrder}{$artifacts, \mathrm{req\_as}, n\_artif, n\_phen$}
     \State $orders \leftarrow \{\{as:\{\},\; ps:\emptyset,\; nums:\{\},\; max:0,\; rest:n\_phen\}\}$
     \Repeat
      \ForAll{$order \in orders \;\; \mathrm{s.t.} \;\; length(order.as)$ is minimum}
       \State $orders \leftarrow orders \setminus \{order\}$
       \ForAll{$a \in (artifacts \setminus order.as)$}
        \State new $neworder$
        \State $neworder.as \leftarrow append(order.as, \{a\})$
        \State $neworder.ps \leftarrow \reqas{neworder.as}$
        \State $neworder.nums \leftarrow append(order.nums, \{|neworder.ps \setminus order.ps|\})$
        \State $neworder.max \leftarrow max(neworder.nums)$
        \State $neworder.rest \leftarrow n\_phen - sum(neworder.nums)$
        \If{$\mathrm{not} \; (\exists o \in orders \;\; \mathrm{s.t.} \;\; \Call{CertainlyBetter}{o, neworder})$}\label{alg:pruning1}
         \State $orders \leftarrow orders \cup \{neworder\}$
         \ForAll{$o \in orders$}
          \If{\Call{CertainlyBetter}{$neworder, o$}}\label{alg:pruning2}
           \State $orders \leftarrow orders \setminus \{o\}$
          \EndIf
         \EndFor
        \EndIf
       \EndFor
      \EndFor
     \Until{$\forall o \in orders \;.\; length(o.as) = n\_artif$}
     \State \Return $orders$
    \EndFunction
    \end{algorithmic}
   \end{center}
  \end{algorithm}

  \begin{table}[t]
   \caption{All Introduction Order of Artifacts in Library Management System}
   \label{033550_10Oct12}
   \begin{center}
    \begin{tabular}[t]{c c c}
     \hline
     $(\introa{i}{A})_{i=1,2,3}$ & $(|\introp{i}{A}|)_{i=1,2,3}$ & Effectiveness Rank \\
     \hline
     $(\{a\}, \{b\}, \{c\})$ & $(6, 1, 1)$ & 1 \\
     $(\{a\}, \{c\}, \{b\})$ & $(6, 2, 0)$ & 2 \\
     $(\{c\}, \{a\}, \{b\})$ & $(6, 2, 0)$ & 2 \\
     $(\{c\}, \{b\}, \{a\})$ & $(6, 2, 0)$ & 2 \\
     $(\{b\}, \{a\}, \{c\})$ & $(7, 0, 1)$ & 3 \\
     $(\{b\}, \{c\}, \{a\})$ & $(7, 1, 0)$ & 3 \\
     \hline
    \end{tabular}
   \end{center}
  \end{table}

 \section{Discussion}
  \label{152306_10Oct12}

  \subsection{Related Work}
   There are some studies on requirement engineering methods for modeling in Event-B.
   In \cite{DEPLOYD19},  Problem Frames and Event-B are applied successfully on an industrial project. The authors constructed a problem diagram before modeling in Event-B. They associated elaborations of phenomena in problem diagram with a data refinement in Event-B. The work of \cite{hallerstede2012method} proposed an iterative process of requirement modeling and validation. The authors connected reasoning about artifacts with refinement in Event-B. In \cite{FM2009-UML-B}, Event-B specifications are derived from class and state-machine diagrams. However, refinement strategy planning is not covered in these studies.

   There have been many studies which aim at deriving formal specification in other methods than Event-B from natural language specifications or diagrams like UML \cite{Miriyala1991,Vadera1994,Macias1995,Jin1997,Ilic2007,Cabral2008,Wang2010,Elbendak2011,VDM-UML} but refinement is not considered in these studies either.

   The authors of \cite{matoussi2011goal} proposed a method to derive an abstract specification of an event. In this study, patterns of correspondences between a KAOS goal model and an event in Event-B are provided. The patterns also consider a part of proof obligations that will be generated. From the point of view of refinement strategy planning, this method transforms a refinement strategy planning for an event into a refinement strategy planning in a KAOS goal model. On the other hand, our method plans refinement strategy of the whole model by considering the constraints on introduction between elements in the system. Thus our approach can be considered as complementary to this work.

  \subsection{Future Work}
   \label{ssec:fw}
   Further refinement of the proposed method is the primary part of the future work.
 
   First, we assumed every artifact is not changed through refinement. However, that is not the case in realistic situations. There are many cases that some artifacts are strengthened in concrete models by using newly introduced phenomena.
 
   Refinement of events is also neglected. In realistic situations, many events are refined through refinements. For example, both event ``Loaning a reserved books'' and ``Loaning a book from the open stack'' can refine an event that only includes ``Add one to loan state'' state transition.
 
   Moreover, as we viewed in Section~\ref{151914_6Oct12}, we assumed that the complexity of modeling can be measured only by the number of phenomena. However this criterion is too rough. For instance, the importance of the number of events and that of variables are different. Thus, finer analysis of complexity of modeling is needed.

 \section{Conclusion}
 \label{134138_11Oct12}
  This paper has aimed at resolving complexities in planning of refinement strategy by considering semantics constraints of Event-B. Refinement strategy planning is an important and difficult phase in modeling in Event-B. Therefore, the proposed method facilitates ordinary users to leverage Event-B. Although much work remains as discussed in Section~\ref{ssec:fw}, we believe this work promotes systematic use of formal specifications, more independently from specific knowledge and skills.

 \bibliographystyle{splncs03.bst}
 \bibliography{tkobayashi_ds-event-b12}
 
\end{document}